\documentclass[a4paper,10pt]{article}
\usepackage{graphicx}
\begin{document}
\newcommand {\be}{\begin{equation}}
\newcommand {\ee}{\end{equation}}
\newcommand {\bea}{\begin{eqnarray}}
\newcommand {\eea}{\end{eqnarray}}
\newcommand {\nn}{\nonumber}

\title{Open theoretical problems in the physics of aperiodic systems}
\author{Anuradha Jagannathan}

\maketitle

\begin{abstract}
Quasicrystals have intrigued and stimulated research in a large
number of disciplines. Mathematicians, physicists, chemists,
metallurgists and materials scientists have found in them a fertile
ground for new insights and discoveries. In the quarter century that
has ensued since the publication of the experimental observation of
a quasiperiodic Al-Mn alloy \cite{shecht}, many different kinds of
quasiperiodic alloys have been manufactured and studied. The
physical properties of quasicrystals are no less interesting than
the unusual structural properties that led to their discovery in
1984. In this review, I present some of the properties that
characterize quasicrystals, 
briefly discuss several types of theories that have been put forward, and
describe some new behaviors that might be investigated by experiment.

\end{abstract}


\section{ What characterizes a quasicrystal ?}
Diffraction patterns are widely used to distinguish different types
of structures in condensed matter physics. The 1984 paper of
Shechtman et al \cite{shecht} entitled "Metallic phase with long ranged
orientational order and no translational symmetry" studied the alloy
using electron and X-ray diffraction. The two main characteristics
of these patterns seemed to contradict each other: a) there were
well-defined spots, indicating a long range order - as in crystals -
in the way atoms were placed, and b) the bright spots in one of the
planes were arranged in a 5-fold symmetric way. Elementary
geometrical arguments rule out, as is well-known, the possibility of
such 5-fold axes in a periodic crystalline array. As better and
better quasicrystals have appeared on the scene, the diffraction
patterns have evolved to eliminate much of the initial diffuse
scattering, showing that these materials have excellent long range
structural order, rivalling those of the best crystals. 

Since those early days, many new experimental
probes have become available, including high resolution
transmission electron microscope (HRTEM), scanning tunneling
microscopy and spectroscopy (STEM) or low energy electron diffraction (LEED),
to name just a few that are commonly used to study surface layers of quasicrystals. Such studies lead in a  natural way
to the notion of atomic clusters. These are used to explain, for
example, the repeating motifs that are seen in the STM or the LEED patterns.
The question of how
such clusters can be put together to form an ordered quasicrystal
then arises, and many authors have addressed this
nontrivial problem ( see for example the compilations in \cite{trebin,fujiwara}).

While the concept of clusters arises when viewing
quasicrystalline surfaces at close range, long range quasiperiodic
order as seen in diffraction studies is perhaps more naturally described in terms of quasiperiodic tilings.
These are the analogs of the ``Bravais lattices" for periodic systems, and give
a good description of the orientational order and positional long
range order seen in quasicrystal diffraction patterns. Tilings are
constructed from a small number of tiles arranged so as to avoid
overlapping or holes. Quasiperiodic tilings usually possess a hierarchically organised structure, where tilings can be "inflated"("deflated") so as to obtain a similar tiling on a larger(smaller) length scale. It was realised early on that quasiperiodic tilings can be obtained by a projection down from
a higher dimensional periodic array. The important early papers can be found in \cite{steinhardt}, and the basic idea is illustrated in Fig.1 below, which indicates schematically how one can obtain quasiperiodic tilings
such as the Fibonacci chain, the Penrose tiling, or the
icosahedral tiling. Points of the $D$ dimensional hypercubic lattice that lie within an infinite strip are projected
onto the irrationally oriented $d<D$
dimensional physical space in which the tiling is defined. The points in the strip are chosen such that their
projections onto the perpendicular subspace
(the remaining $D-d$ perpendicular directions) fall inside a
polyhedral volume, called the selection window. Thus, {\it every vertex of
the tiling possesses a counterpart 
in the compact perpendicular space window}. This property provides one with a convenient way of representing certain spatial properties of quasicrystals, as we will see in our magnetic model later.

The method of projection manifestly implies long range order, admittedly of a complex type.
It also leads to repetition of local atomic clusters
(obtained by decorating the vertices of the $D$ dimensional
lattice). The method can be generalized to obtain many types of
structures, including various kinds of disordered ones. In addition, by
increasing the dimension $D$, one can go from the crystal ($D=d$) to
more and more strongly aperiodic structures. This provides
interesting models, for example, when one wishes to study the
localization of electrons, or, as we consider here, magnons in different quasiperiodic structures.

\begin{figure}[ht]
\begin{center}
\includegraphics[scale=0.350]{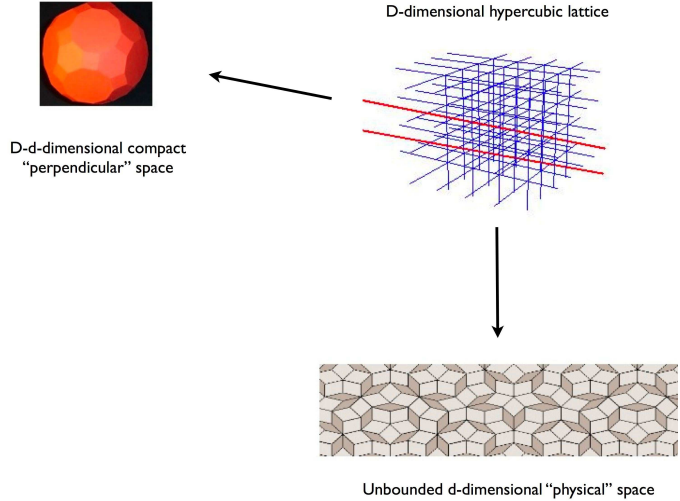}
\vspace{.2cm} \caption {Schematic view of the cut-and-projection method: all the vertices of a
D-dimensional lattice (center) in the region delineated by irrational planes are chosen for projection.
The two possible projections give (bottom) vertices of a tiling in the physical space, or (left) a densely filled
region of the complementary ``perpendicular" space} \label{project.fig}
\end{center}
\end{figure}
\section{Electronic, magnetic and vibrational properties }

Leaving the subject of determination of the structures of specific quasicrystals, we now raise
the question: can one make some general statements about the consequences of quasiperiodic structural order
on physical properties of these materials ? 

A common feature of most quasicrystalline alloys is the
dip, or pseudogap, in the electronic density of states close to the Fermi level
\cite{pseudogap}. This pseudogap presumably leads, following the argument of
Hume-Rothery \cite{dubois}, to the formation of the quasicrystalline
phase for certain values of the electron concentration. The
pseudogap is found in
approximants as well. In other words, it depends primarily on the short range
structural details, with modifications due to the long range order. Short range structural order also determines the magnitude of
the electrical conductivity to a large extent. It plays an important role in magnetic properties of magnetic quasicrystalline alloys
such as the Curie-Weiss temperature, or the spin glass transition
temperature. 

To observe properties arising due to long range quasiperiodic order,
one has to go to low temperatures and of course make measurements on
systems which are perfectly quasiperiodically ordered on a
sufficiently long length scale. Some phenomenological models have been proposed for the transport in quasicrystals, such as a prediction 
based on hierarchical structure models for the
temperature dependence of the electrical conductivity of icosahedral
alloys \cite{janot}. Vibrational properties, similarly, should reflect the hierarchical structure present in quasicrystals. At low energies, phonon modes may for example show anomalous dispersion laws. This in turn could lead to a nontrivial 
temperature dependence of the lattice contribution to the thermal
conductivity. In the case of aerogels which are disordered fractals, such effects have been shown in neutron scattering and by thermal conductivity experiments. Recently studies have been carried out to investigate the thermal conductivity in a number of quasicrystals \cite{smont}.

Computational approaches such as ab-initio and molecular dynamics calculations have been successful in
fitting some experimentally observed electronic \cite{krajci} and vibrational
\cite{deboiss} properties. Analytical results are
on the other hand scarce, even in the simplest of models, due to the lack of
appropriate theoretical tools. It is however important to study simplified models for their ability to isolate some simple generic
behaviors. Thus, for example, tight-binding models are useful for
systems such as graphene, to mention one much studied
current problem, or high-temperature superconductors. Such models allow to predict behaviors independent (to a first approximation) of details of chemistry and could be applied to a family of related compounds.

In this context, a promising direction of research, in terms of
isolating the physics of quasiperiodicity, is the attempt to
generate a simple single-element quasicrystal by deposition on an
appropriate substrate. This is described in several contributions in this
workshop \cite{workshop}. As will be discussed in more detail
in the last section, surface probes on such quasicrystalline films
should provide direct evidence of some of the most interesting
properties of quasiperiodic materials. One interesting challenge
would be to excite and observed critical states,
somewhat along the lines of the ``quantum corral" experiment on a
copper substrate \cite{ibm} in which it was possible to image a
electron state localized on Fe atoms that were dragged into a
ring-shape by an STM tip. The electron problem is unfortunately complicated by electron-electron interactions, but magnon modes on a quasiperiodic surface should exhibit this type of critical states, as we discuss next.

The following section describes a problem that
was studied recently for two dimensional quasicrystals: the antiferromagnetic ground state, and magnon excitations. 
We will see that many properties observed in the
phonon and electron problems are carried over to the magnetic case,
showing that quasiperiodic antiferromagnets could provide a new
testing ground for unusual properties of quasicrystals.

\section{Model of a quasiperiodic quantum antiferromagnet}
We illustrate the properties of a Heisenberg antiferromagnet in a
2 dimensional quasiperiodic structures such as the Penrose rhombus tiling or the Ammann-Beenker (octagonal) tiling. The
Hamiltonian is taken to be of the simplest possible form:

\be H = J\sum_{\langle i,j\rangle} \vec{S}_i.\vec{S}_j \ee

where the spins are situated on vertices of the tiling, and the
bonds act along edges connected nearest neighbors. The bonds are all assumed of the same
magnitude, $J>0$, and the most interesting situation corresponds to
the case of $S=1/2$ spins. The questions we address
are a) what is the quantum ground state of this system ? and b) what
are the equivalent of magnons (the elementary excitations in a
periodic antiferromagnet) in a quasiperiodic material ?

\begin{figure}[ht]
\begin{center}
\includegraphics[scale=0.450]{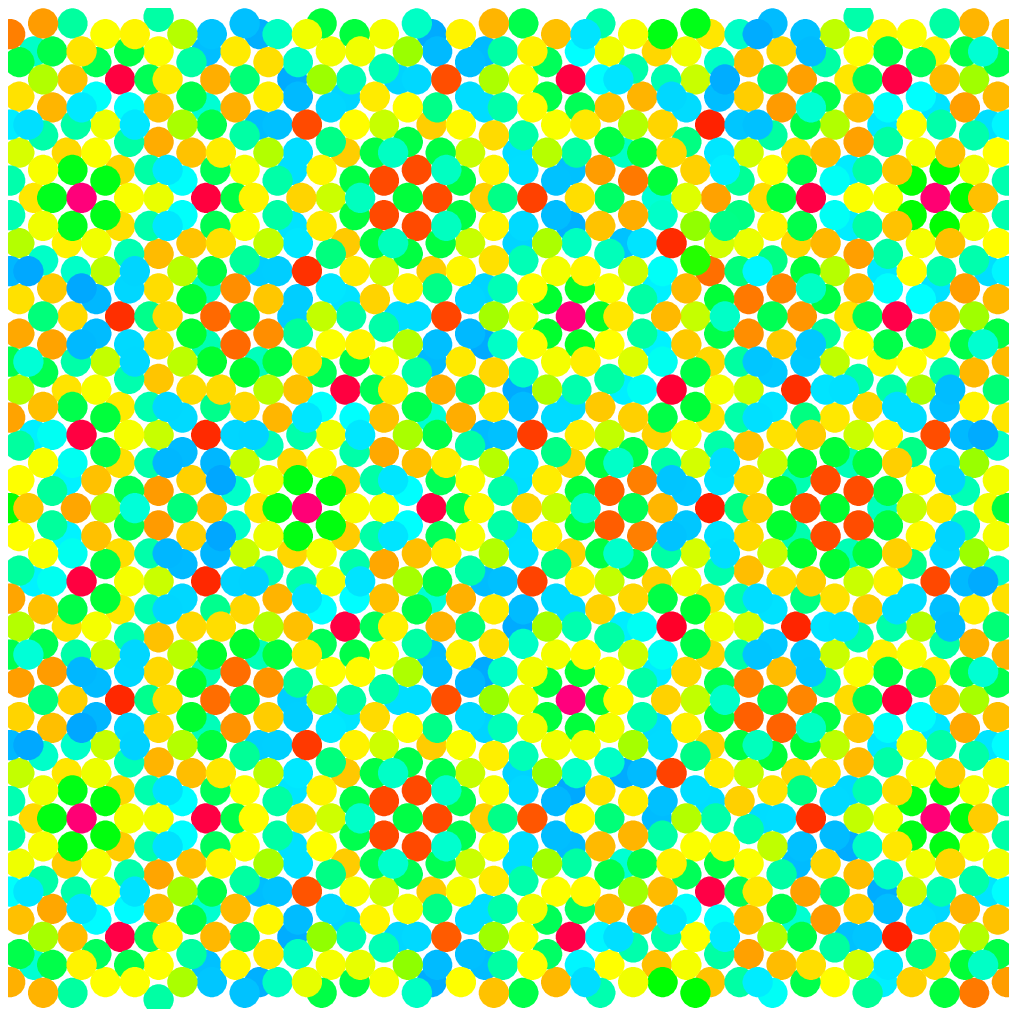} \hfill
\includegraphics[scale=0.35]{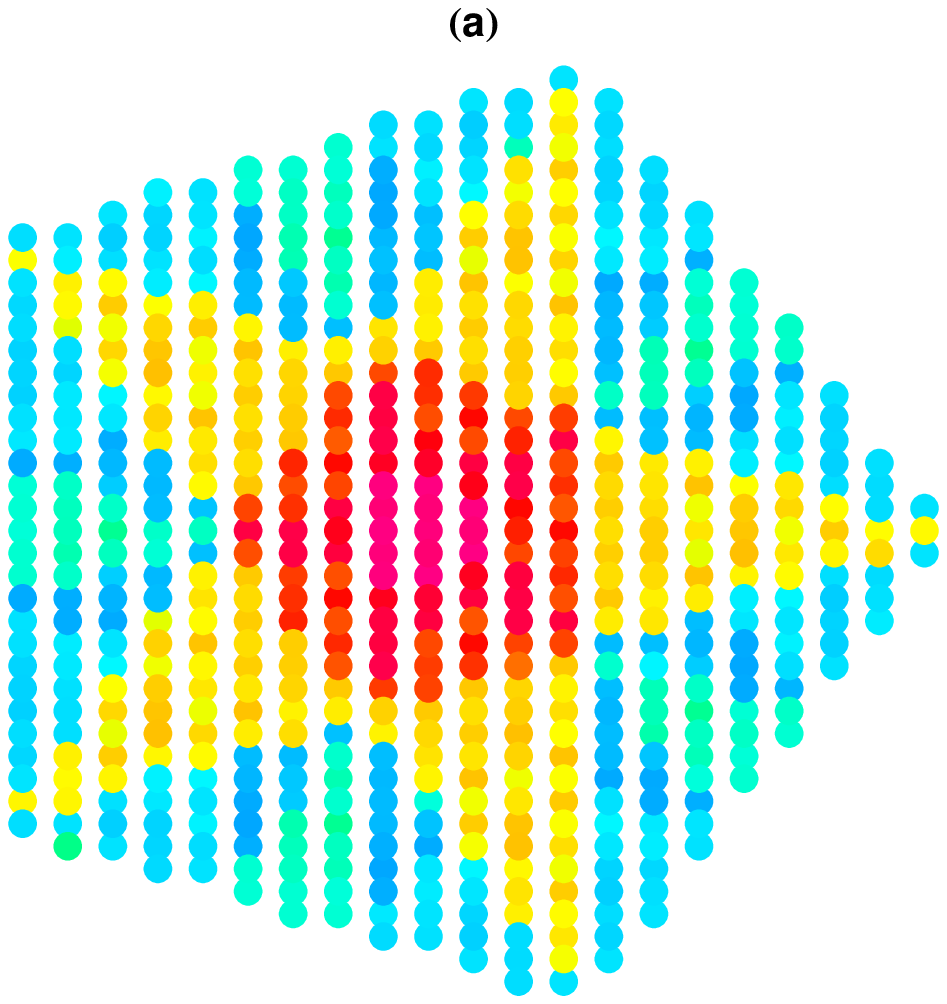} 
\includegraphics[scale=0.35]{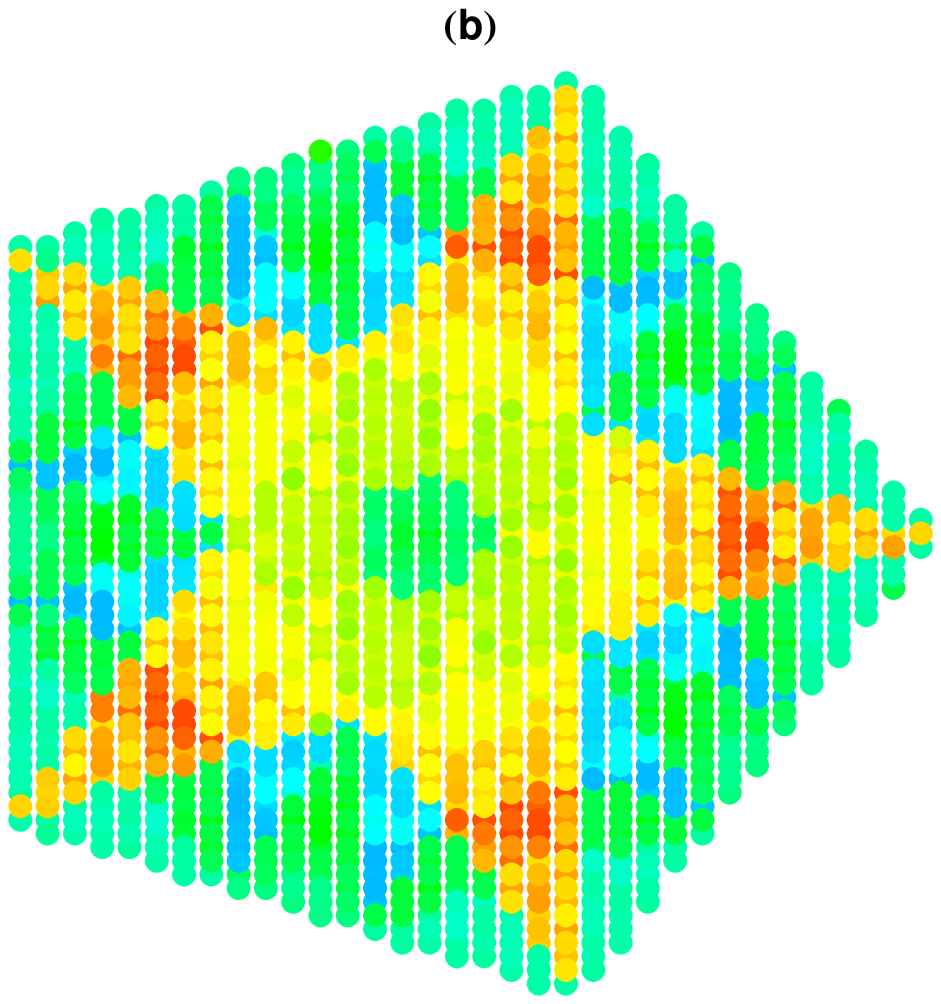} \vspace{.2cm}
\vspace{.2cm} \caption {Color representation of the distribution of absolute values of the local magnetizations in the
ground state of a Penrose antiferromagnet. (Left) as seen in real space.
(Center and Right) after projection onto two regions of the perpendicular space (colors range from red for the smallest values, to blue for the
highest values)} \label{fig:gsmag}
\end{center}
\end{figure}

The answers were found by approximate methods (both numerical, using a spin wave
expansion and analytical, using renormalization group) and by Quantum Monte Carlo calculations (see \cite{penroseafm} for details and references).
The spin wave method, analogous to the one used for periodic systems,
does not require wave properties, as one can work in a real space
basis without need for any Fourier transforms. In agreement with the QMC results, we find the result shown in Fig.\ref{fig:gsmag} for the ground state magnetic
configuration of the spins in one of the approximants of the Penrose tiling. In the figures, each site has been colored according to the absolute
value of its local magnetization $\vert \langle S_i\rangle\vert$. The left hand figure shows the sites in real space (the tile edges have not been drawn), showing that the ground state magnetization is quite complex. One can
discern a self-similar structure, even within the small region that is
shown. In the center and right figures, the same ground state is represented using the perpendicular
space coordinates of the sites of the tiling (a) and b) indicate two out of the four domains that constitute the full set of vertices). These figures show an alternative view of the ground state structure, which is perhaps simpler, with colors grouped together in domains. This
representation is a compact way to show the ground state magnetic
structure, and also provides an elegant way to demonstrate
self-similarity (not discussed here).

The ground state shown above is the result of quantum fluctuations -- magnons, or spin excitations -- of the quasiperiodic Heisenberg model. The magnon spectrum and the real space-dependence of the magnons have been
determined in spin wave theory. The density of magnon states is very reminiscent of that of scalar 
phonons in quasiperiodic tilings \cite{los2}, with a linear low energy "dispersion relation" for low energy modes, as in other types of solids.
The values of the effective spin wave velocity depend of course on the specific structure. One can compare the values of the velocity with that found in the square lattice, which is a reference system having the same average connectivity (z=4) as the Penrose tiling or the octagonal tiling. One finds that the velocities decrease in the order square lattice $\rightarrow$ octagonal tiling $\rightarrow$ Penrose tiling, reflecting the rise in complexity as one goes from one structure to the next. When the quasiperiodic tiling is disordered by permuting tiles (adding frozen phason disorder), the sound velocity $increases$, whereas experience based on periodic systems would indicate the contrary \cite{szallas}. This is the equivalent of a phenomenon already observed in studies of electron motion in quasiperiodic environment, as was shown for example by Passaro et al \cite{passa}, who showed that the diffusion exponent for spreading of a wavepacket increased in a disordered two dimensional tiling compared to the nonrandom tiling. 
  
\begin{figure}[t]
\begin{center}
\includegraphics[scale=0.45]{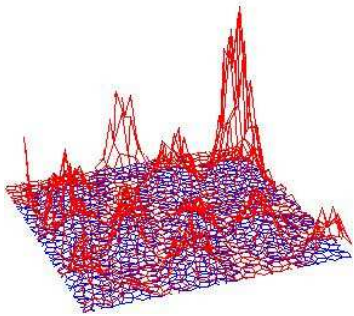} \vspace{.2cm}
\caption{(Color online) A representation of a magnon wavefunction in a Penrose tiling, corresponding to a low energy state $E=0.82$, showing the square of the wavefunction for each of the sites in a finite region of the tiling. Although the system size is small, it illustrates the recurrent peak
structure and power law decay.} \label{fig:critwave}
\end{center}
\end{figure}

Other interesting properties of the magnetic model are related to the spatial behavior of the spin excitations. The magnons have an extended
character at low energies, with possibly critical states, as shown
in Fig.\ref{fig:critwave}. The wavefunction for the energy
eigenstate shown has a central peak, and lesser peaks at sites
situated at large distances. This type of wavefunction leads to
spin-spin correlations that are very large for different subsets of
sites, depending on the energy \cite{icq10}. Such effects would be observable for quasiperiodic magnetic layer of atoms, a difficult but in principle reachable experimental goal! The results obtained for our spin model lead us to speculate that nonlocal effects and quantum
entanglement features should be particularly interesting in these systems. 

\section{Discussion and Conclusions}
Where quasicrystals are concerned, theory has lagged behind experiment in terms of explaining the observed electrical, thermal and magnetic properties. This is due to the fact that the available theoretical tools developed for periodic crystals are either not valid or else not easily applicable in these aperiodic materials. Simulations of quasicrystals are computationally intensive, and hence results are obtained for rather small system sizes, depending on the available memory and time constraints. On the other hand, simple models such as those outlined in this paper predict that quasicrystals may possess highly unusual and distinctive physical properties that are not shared by crystalline or amorphous materials. The experimental realization of such simplified models is difficult but not impossible, in view of the recent achievements in the deposition of atoms on quasiperiodic surfaces to form two dimensional structures. Further theoretical and experimental progress is clearly needed in order to understand the properties of this class of materials.

\end{document}